\documentclass[a4paper,superscriptaddress,twocolumn,prl]{revtex4}
\usepackage[pdftex]{graphicx}
\usepackage[]{natbib}
\pdfoutput=1
\newcommand{\sign}{\mathop{\mathrm{sign}}}

\begin{document}

\title{Optically guided linear Mach Zehnder atom interferometer}

\author{G. D. McDonald}
\email{gordon.mcdonald@anu.edu.au}
\homepage{http://atomlaser.anu.edu.au/}
\author{H. Keal}
\author{P. A. Altin}
\author{J. E. Debs}
\author{S. Bennetts}
\author{C. C. N. Kuhn}
\author{K. S. Hardman}
\author{M. T. Johnsson}
\author{J. D. Close}
\author{N. P. Robins}

\affiliation{Quantum Sensors Lab, Department of Quantum Science, Australian National University, Canberra, 0200, Australia}

\date{\today} 

\begin{abstract}
We demonstrate a horizontal, linearly guided Mach Zehnder atom interferometer in an optical waveguide. Intended as a proof-of-principle experiment, the interferometer utilises a Bose-Einstein condensate in the magnetically insensitive $\left| F=1,m_{F}=0\right\rangle$ state of Rubidium-87 as an acceleration sensitive test mass. We achieve a modest sensitivity to acceleration of $\Delta a = 7\times10^{-4}$m/s$^{2}$. Our fringe visibility is as high as 38\% in this optically guided atom interferometer. We observe a time-of-flight in the waveguide of over half a second, demonstrating the utility of our optical guide for future sensors.
\end{abstract}

\maketitle


Over the past decade there has been significant interest in the application of Bose-Einstein condensates (BEC) to the development of compact inertial sensors based on magnetically guided ultra-cold atoms \cite{RevModPhys.79.235, Farkas}.   Trapped atom systems offer the possibility of the ultra-high precision sensing demonstrated by free-space atom interferometry \cite{Peters_Mobile,quntas} in a more compact package.  Atoms can now be Bose-condensed \cite{Schneider,Ott,Hansel,NakagawaFastChipBEC}, guided \cite{Key,Leanhardt02}, split \cite{Denschlag99,Cassettari,Hinds}, switched \cite{Muller01}, recombined \cite{tonyushkin:094904} and imaged \cite{huet:121114,Smith:11} in reconfigurable magnetic potentials which support the atoms against gravity. 
Typical geometries for magnetically trapped atom interferometers use either atoms bound to a trap which is adiabatically deformed \cite{Shin04,Shin05,Jo,Schumm} or a magnetic guide in which atoms are manipulated using a standing wave \cite{Garcia,PhysRevA.70.013409,Wang,Wu,Su}. 

Precision in these schemes is usually limited by both the roughness of the magnetic waveguide potential which causes decoherence and fragmentation of the condensate \cite{Schumm05,Jones, Leanhardt03, Fortagh}, as well as interaction induced dephasing due to the tight trapping potentials used in magnetic guiding \cite{Chen,Horikoshi,Kreutzmann}. 
Methods used to address these problems have included a Michelson configuration which is only sensitive to relative acceleration between the two arms \cite{Wang,Kafle}, a constant displacement scheme with an inherently reduced scaling in sensitivity to absolute accereration \cite{Su}, or trapping currents oscillating in the kHz range which smooths the potential but causes unwanted heating \cite{Trebbia07,Bouchoule}. The impact of these problems has been highlighted in Ref. \cite{Marti}.


An alternative solution using optical trapping and manipulation of ultra cold atoms has the advantage of being inherently smooth. Optical elements have been constructed which guide \cite{Bongs,Guerin,Gattobigio09,Dall}, reflect \cite{Fabre,Cheiney} and split \cite{Houde,Gattobigio12,Dumke} atom clouds. Recently, a ring interferometer has been constructed to measure rotation \cite{Marti}. Additionally, relatively large BECs can be quickly produced in optical traps ($10^5$ atoms in 500ms \cite{Clement}) and the atoms in an optical trap can be confined in any internal state, allowing the trapping of magnetically insensitive ensembles \cite{AltinNJP}.

 
 \begin{figure}
\centering{}
 \includegraphics[width=1\columnwidth]{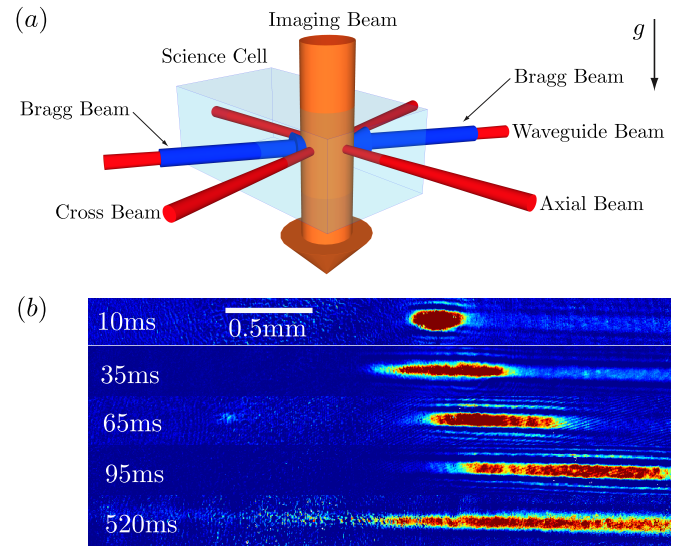}
 \caption{(color online) (a) The geometry of our optically guided atom interferometer. A BEC is formed in an optical dipole Ôtriple trapÕ at the intersection of three far-detuned beams. Two of these are switched off to release the atoms into the third beam, the waveguide.  A MZ atom interferometer is constructed using Bragg transitions from counter-propagating beams aligned along the waveguide.  We image the resulting momentum states using a vertical absorption imaging system. A second absorption imaging system, not shown in this diagram, has its axis in the horizontal plane between the cross and waveguide dipole beams. 
 (b) Images showing expansion of the condensate in the waveguide after different expansion times. Because gravity slowly pulls the atoms out of the field of view of our imaging system, the image after 520ms expansion is of a condensate thrown `up hill' by a 6$\hbar k$ Bloch acceleration, and then allowed to fall back into the field of view. }
 \label{science cell}
 \end{figure}


In this paper we present the first linear, optically guided atom interferometer in an inertially sensitive configuration. A BEC of $^{87}$Rb is loaded into an atomic waveguide constructed from a far-detuned optical dipole beam (Fig.  \ref{science cell}). The atoms are then transferred into the first-order magnetically insensitive $\left| F=1,m_{F}=0\right\rangle$ spin state. A Mach-Zehnder (MZ) atom interferometer with $4\hbar k$ momentum splitting is constructed using counter-propagating Bragg beams aligned co-linear with the waveguide. The phase $\Phi$ of a MZ atom interferometer is given by \cite{MullerBragg}
\begin{equation}
	\Phi=n(2\mathbf{k\cdot a}-\alpha)T^2+n(\phi_1-2\phi_2+\phi_3)
	\label{eq:IntPhase}
\end{equation}
 where $\mathbf{k}$ is the wavevector of the light used in the $n$th order Bragg transitions, $\mathbf{a}$ is the acceleration experienced by the atoms from external forces, $\alpha$ is the rate at which the angular frequency difference between the Bragg beams is swept,  $T$ is the time between pulses in the interferometer of total length $2T$ and $\phi_j$ is the phase of the $j$th Bragg laser pulse. Tuning the interferometer phase $\Phi$ to zero using $\alpha$ provides a measure of the acceleration along $\mathbf{k}$. We demonstrate this by measuring the small residual component of gravity along the near-horizontal waveguide.


\begin{figure}
\centering{}
 \includegraphics[width=1\columnwidth]{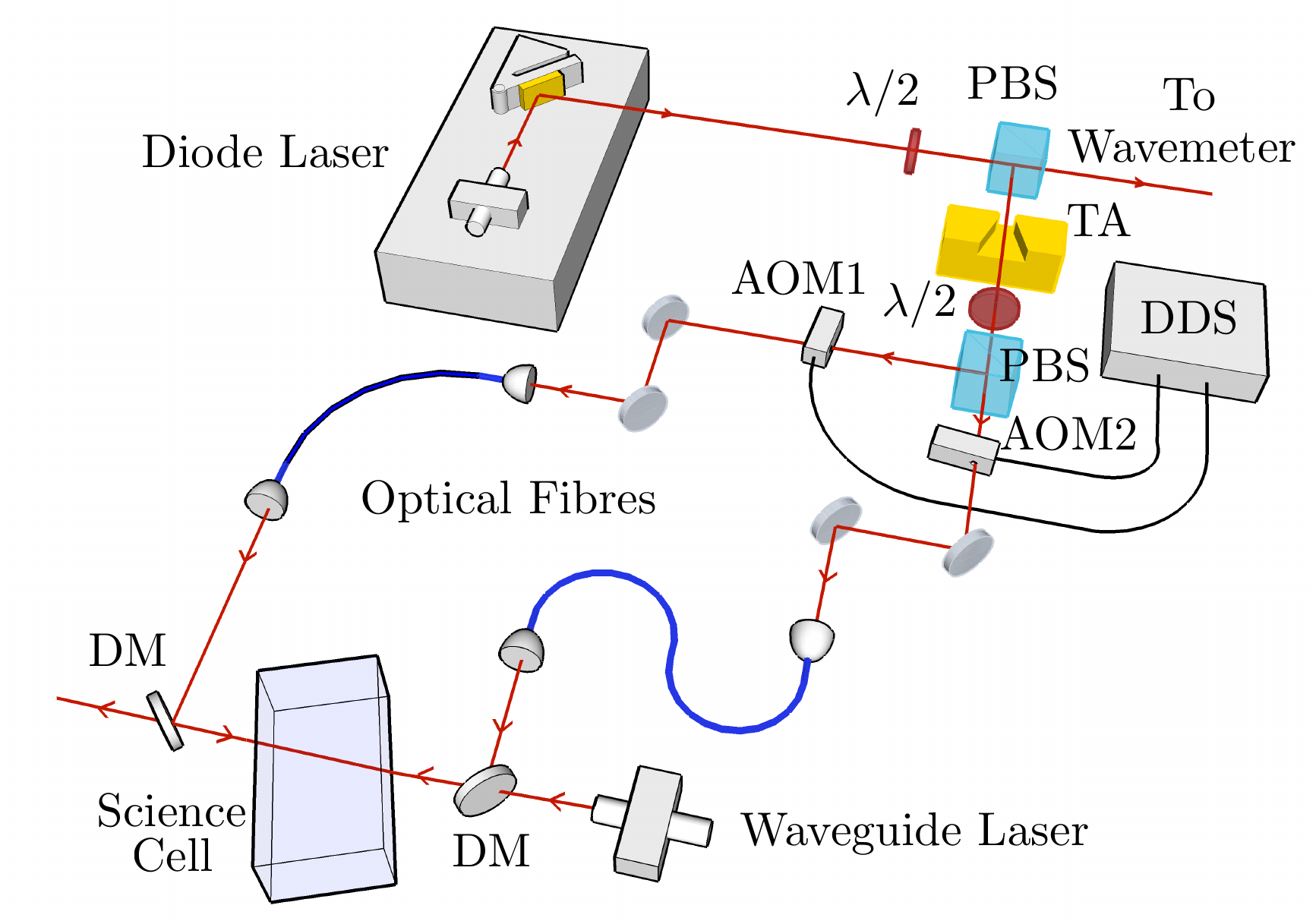}
 \caption{(color online) Our Bragg laser system consists of two counter-propagating 780nm beams aligned co-linear with the waveguide and detuned from one another on the order of tens of kHz. The beam from an external cavity diode laser detuned by $\sim130$GHz from the $D_2$ line of $^{87}$Rb (as measured using a HighFinesse WS2 Wavemeter) is used to seed a tapered amplifier (TA). The output from the TA is split between two acousto-optic modulators (AOM) by a polarising beam splitter (PBS) with a half-wave plate ($\lambda/2$) for frequency and amplitude control. Each AOM is driven near 80MHz by one of two amplified, phase-locked channels from a direct digital synthesiser (DDS, Spincore PulseBlaster). The modulated beams are coupled into separate optical fibres which bring the beams near to the atoms. Dichroic mirrors (DM) are then used to align these Bragg beams counter-propagating and co-linear with the waveguide.}
 \label{BraggFigure}
 \end{figure}
 		

We produce $^{87}$Rb condensates using the machine described in Ref \cite{Altin}. Briefly, we evaporatively cool atoms in their $\left|F=1, m_F=-1\right\rangle$ lower ground state in a quadrupole-Ioffe configuration magnetic trap before transferring them into an optical `triple trap'. The `triple trap' is constructed using three red-detuned dipole beams (see Fig. \ref{science cell}). The cross and axial beams are sourced from a single laser (SPI RedPower compact) operating at $1090$nm, while the third beam (SPI RedPower HS) which operates at $1065$nm is also later used as our optical waveguide. The $1/e^2$ waist radii of our axial, cross and waveguide beams are measured to be $135\mu$m, $135\mu$m and $80\mu$m respectively. The waveguide beam is held on at a constant power of $4.5$W. The crossed dipole beams are adiabatically ramped down from $4.5$W to $1.65$W over $1.5$s which further evaporatively cools the atoms, producing a BEC of $5\times10^{5}$ atoms. Our slow repetition rate of 0.5/min is largely dominated by the need for thermal dissipation from our magnetic trap, and it is possible to form BEC much faster than this \cite{Lin,Clement}.

     \begin{figure*}
\centering{}
 \includegraphics[width=1\textwidth]{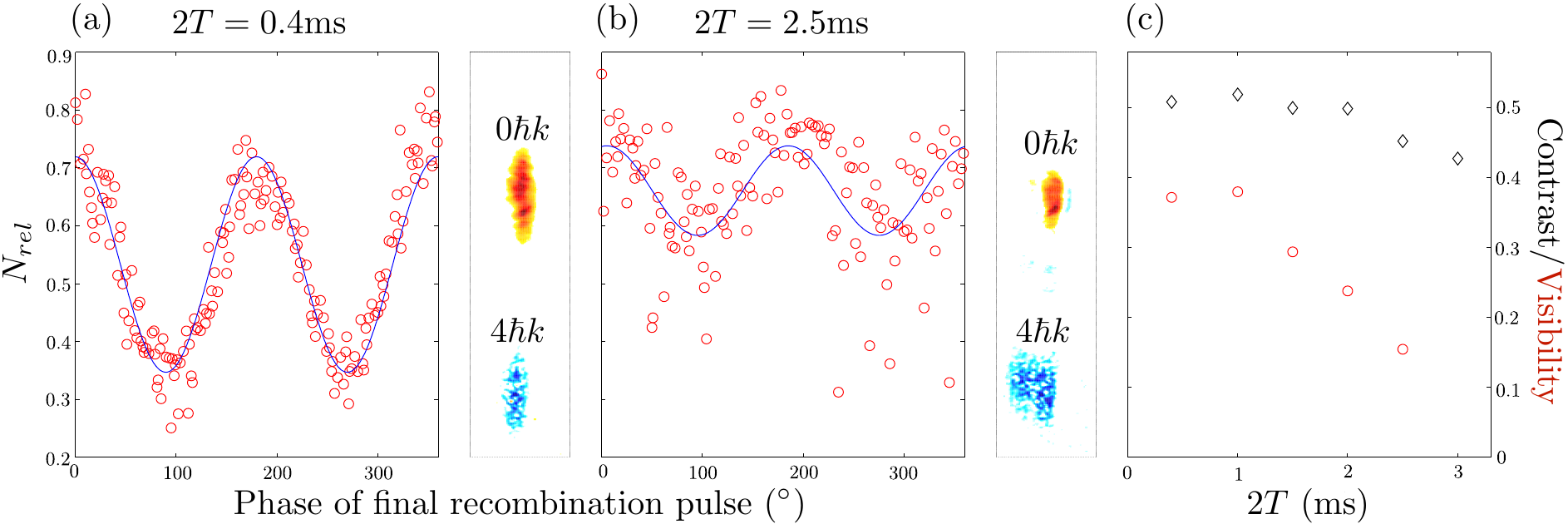}
 \caption{(color online) We obtained fringes in Mach-Zehnder configuration with $4\hbar k$ momentum splitting. Measured fringes (red circles) and a sinusoidal fit (blue line) of the form $N_{rel}=A\cos(2\phi_3+\Phi)+c$ for (a) $2T=400\mu$s and (b) $2T=2.5$ms. The density plot next to each fringe is a Fourier component of our absorption images for all recombination phases $\phi_3$ (see text), and shows the sections of our absorption images which contribute to each state of the interferometer. The $0\hbar k$ (red atom cloud) and $4\hbar k$ state (blue atom cloud) are separated by 870$\mu$m.  (c) Visibility (red circles), $2A$, as measured by the sinusoidal fit to each fringe set. Contrast (black diamonds) as measured by range of data $N_{rel}$ from the 2nd percentile to the 98th percentile, is shown for comparison to indicate possible gains in fringe visibility after the elimination of phase noise.}
 \label{fringes}
 \end{figure*}

  
To release the atoms into the waveguide we ramp the crossed dipole beams down to $70$mW over 0.5s before switching them off entirely. The remaining optical waveguide beam has transverse and axial frequencies of 114Hz (measured by exciting a trap oscillation) and 1Hz (calculated from the beam properties) respectively, and is on a tilt of less than $1^{\circ}$ with respect to gravity. Consequently the atoms slowly accelerate out of the field of view of our vertical imaging system ($\approx$ 3mm) after around 100ms. We observed the condensate expanding along the waveguide without aberration for times on the order of 0.5s (Fig. \ref{science cell}) by using a $6\hbar k$ Bloch acceleration \cite{Clade} up the slight incline and observing the atom cloud as it falls back down the waveguide. 
After the BEC is released into the waveguide, we allow it to expand axially for 20ms to reduce any mean-field effects which may be present due to inter-particle interactions at higher density \cite{Debs}. After expansion we measure the momentum width in the directions axial and transverse to the waveguide to be $0.8 \hbar k$ and $0.2\hbar k$ respectively. Using time of flight observations we have determined that the majority of the atoms occupy the transverse ground state of the waveguide.

 While the BEC expands along the waveguide a constant magnetic field of $30$ Gauss is applied by a pair of Helmholtz coils to define the spin axis. During this time the atoms are transferred into the first-order magnetically insensitive $\left|m_{F}=0\right\rangle$ state using a Landau-Zener radio frequency sweep. We can verify that the atoms are in the $\left|m_{F}=0\right\rangle$ state by hitting the cloud with a short magnetic pulse, knocking them out of the waveguide if they are in the $\left|m_{F}=-1\right\rangle$ state but leaving them trapped if they are in the $\left|m_{F}=0\right\rangle$ state.

 We use Bragg transitions to coherently split, reflect and recombine our atomic wavepacket in momentum along the waveguide \cite{Debs,MullerBragg}. Our Bragg setup is shown schematically in Figure \ref{BraggFigure}. For counter-propagating beams an $n$th order Bragg pulse, imparting $2n\hbar k$ momentum to the kicked atoms, has a resonance condition given by $\Delta f=n\hbar k^2/m\pi$, where $k$ is the wavenumber of the light and $m$ is the mass of the atoms. We use $\Delta f=30.3$kHz to effect second order Bragg transitions. To account for the doppler shift induced by the acceleration of approximately 0.10m/s$^2$ down the waveguide due to gravity (as measured by time-of-flight in the waveguide), one of the beams is swept by  $\alpha=2\pi\times258$Hz/ms in the laboratory frame so as to remain resonant, with no doppler shift in the frame of the atoms. We use gaussian pulses to achieve optimal momentum state coupling efficiencies \cite{Szigeti,MullerBeamSplitter}.
 
Using the Bragg setup we build a Mach-Zehnder interferometer. First a $\pi/2$ pulse is applied to coherently split the atoms into two momentum states, one initially stationary at $0\hbar k$, the other travelling at $4\hbar k$. After a time $T$ we apply a $\pi$ pulse to invert the two momentum states. After another period $T$, the two halves of the atomic wave packet are overlapped again and we apply a second $\pi/2$ pulse to interfere the two states. We allow these final states to separate along the waveguide for $(35-2T)$ms, then switch off the waveguide to allow ballistic expansion for 5ms to avoid lensing of the imaging light by the narrow, dense cloud of atoms.  Using absorption imaging we count the number of atoms in each momentum state. To remove the effect of run-to-run fluctuations in total atom number, we look at the relative atom number in the $0\hbar k$ state $N_{rel}=N_{0\hbar k}/(N_{0\hbar k}+N_{4\hbar k})$. By scanning the relative phase $\phi_3$ of the final $\pi/2$ pulse, we obtain fringes in $N_{rel}$, and these are shown in Fig. \ref{fringes}.

A simple method to count the atoms in each state is to draw a box around the area where each state is expected and count the atoms in each box for each phase $\phi_3$. To avoid counting non-contributing pixels in our image, which would add unnecessary noise, we use a Fourier phase decomposition algorithm to select which pixels we attribute to each momentum state. For each pixel $i$ in our absorption image we calculate the number of atoms it contains as a function of recombination phase, $n_i(\phi_3)$. We then take the inner product with sinusoids of the expected frequency 
 
 \begin{equation}
\begin{array}{l}
\displaystyle \alpha_i=\int_{0}^{2\pi}n_i(\phi_3)\cdot \sin(m\phi_3)d\phi_3  \\
\displaystyle \beta_i=\int_{0}^{2\pi}n_i(\phi_3)\cdot \cos(m\phi_3)d\phi_3
\end{array} 
\label{eq:fourier coefficients}
\end{equation}
 where $m$ is 2 for a $4\hbar k$ transition. Any oscillatory signal in $n_i(\phi_3)$ of the correct frequency such as $n_i(\phi_3)=A_i\cos(m\phi_3+\Phi_i)$ can be extracted by the relations
  
\begin{equation}
\begin{array}{l}
\displaystyle A_i=2\sqrt{\alpha^2+\beta^2}  \\
\displaystyle \Phi_i=\tan^{-1}(\frac{\alpha_i}{\beta_i})
\end{array} 
\label{eq:xdef}
\end{equation}

For a small phase offset ($\Phi_i\approx0$ for the 0$\hbar k$ state) it is sufficient to simply plot $\beta_i$, as $\left|\beta_i\right|\approx A_i$ and $\sign(\beta_i)\approx\cos(\Phi_i)$, and this has been done in Fig. \ref{fringes}. Ideally, two identifiable components will be visible in an image, the $0\hbar k$ momentum state with $\Phi\approx0$ (with positive amplitude, shown in red) and the $4\hbar k$ momentum state with $\Phi\approx\pi$ (negative amplitude, blue). From this image we select which pixels to include in our regular counting of $N_{0\hbar k}$ and $N_{4\hbar k}$ for all $\phi_3$ by setting a tolerance on $\beta_i$. The optimal tolerance will depend upon the background noise in the image.

 An example of the obtained fringes are shown in Fig. \ref{fringes}. We obtain a visibility of $38\%$ at $2T=1$ms and  $15\%$ at $2T=2.5$ms. By $2T=3$ms, phase noise effectively randomises the final phase of the interferometer, but interference is still visible. Even at $2T=7$ms we still have interference with contrast of $\approx37\%$, albeit with random phase. The phase instability observed at longer interferometer times is likely due to acoustic vibrations affecting the optical fibre out-couplers which bring the Bragg beams to the table. A simple analysis shows that a small fluctuation in the distance $\Delta L$ between fibre out-couplers creates a laser phase offset (in radians) of $\Delta \phi_i=4\pi n\Delta L/\lambda$. For the sake of argument, assume $\Delta\phi_{1,2}=0$, $\Delta\phi_3=\pi/2$ is enough to mask a usable signal, this means that $\Delta L\approx50$nm is enough displacement during the interrogation time $T$ to completely wash out any fringes. This could be caused by a vibration with a 70nm amplitude and frequency around $f=1/3T\approx170$Hz with $2T$=4ms, for example. Indeed, by looking at the beat between our Bragg beams on a low-frequency spectrum analyser we see a significant noise peak between 130Hz and 200Hz in our laboratory. 

The highest sensitivity to acceleration along the guide that we can currently obtain is $\Delta a = 7\times10^{-4}$m/s$^{2}$ at $2T=2.5$ms over 136 runs ($9\times10^{-2}/ \sqrt{Hz}$), and we obtain an acceleration of $a=0.0997(7)$m/s$^2$. For comparison, a free space gravimeter run in the same lab \cite{Debs} had an acceleration sensitivity of $5\times10^{-4}$m/s$^{2}$ at $2T=6$ms over 30 runs ($3\times10^{-2}/\sqrt{Hz}$).   The similar results obtained for both the free space and guided interferometer indicate that it is likely that by vibrationally isolating the sensor and Bragg laser system from the mechanical noise present in our laboratory we can achieve significantly higher sensitivity.  Indeed, a precision atom interferometer based gravimeter, operated in a vibrationally isolated laboratory next to the one in which the current apparatus resides achieves an acceleration sensitivity of $\Delta g\sim 3\times10^{-7}/\sqrt{Hz}$ \cite{the_archive} for $2T=200$ms. The fundamental atomic projection noise limit on acceleration sensitivity for this type of system is given by $\Delta a=1/\sqrt{N}kT^2$ where $N$ is the total number of atoms involved in several runs of the experiment \cite{AltinNJP}. For our longest waveguide propagation time of $2T=520$ms this limit is an enticing $\Delta a=4\times10^{-11}$m/s$^{2}$ $(2\times10^{-9}/\sqrt{Hz})$. In this hypothetical interferometer we would have a maximum displacement between the atom clouds of 3.6mm, or 10\% of the Rayleigh length in either direction and the resulting change in waveguide intensity experienced by the atoms will be less than 1\%.  

There are numerous avenues for future research in this system.  If vibrational noise can be reduced, we can begin to explore the fundamental limitations of signal to noise in the waveguide interferometer, and additionally make a direct comparison to a free space system in the same machine.  The ability to hold all magnetic substates in the same waveguide spatial mode with an arbitrary, constant magnetic field offers another interesting prospect: completely removing the self-interaction in such a system by setting the scattering length to zero  \cite{florence,innsbruck}. In fact, our apparatus can also produce BEC of $^{85}$Rb and manipulate the s-wave scattering length via an easily accessible Feshbach resonance at 155G \cite{Altin}. Combining the optical waveguide interferometer with a time varying scattering length could also allow investigation of squeezing enhanced interferometry \cite{Mattias,Haine,Esteve:2008fk}.   Finally, we have made preliminary investigations of an alternative to two-photon beam splitters and mirrors in the waveguide.  By replacing the Bragg mirror with a blue detuned light sheet at 532nm we have constructed a hybrid interferometer, which will be the subject of an upcoming paper. The system also offers the possibility of superimposing multidimensional lattices onto the propagating atoms to create the equivalent of photonic crystals for the propagating atoms.  

In summary we have demonstrated a proof-of-principle acceleration sensor based upon Bragg interferometry in an optical waveguide. Our Mach-Zender configuration atom interferometer is sensitive to acceleration along the waveguide axis. As the atoms are optically trapped we are able to operate the interferometer with atoms in the magnetically insensitive $\left|F=1,m_{F}=0\right\rangle$ internal state. We have demonstrated clean propagation in the optical waveguide without fragmentation for more than half a second.  In the future, this single axis system could be readily adapted to produce a multi-axis inertial sensor by including two additional orthogonal waveguide atom interferometers.

\end{document}